
\magnification=1200
\baselineskip=20pt
\nopagenumbers
\settabs 20 \columns
\+&&&&&&&&&&&&&&& UM-TH-95-18 \cr
\+&&&&&&&&&&&&&&& hep-ph/9507207 \cr
\+&&&&&&&&&&&&&&& June , 1995 \cr

\vskip2truecm

\centerline{\bf A Finite Group Analysis of the Quark Mass Matrices and
Flavor Dynamics }

\vskip1.5truecm

\centerline {York-Peng Yao}
\medskip

\centerline { \sl Randall Laboratory of Physics, University of
Michigan }
\smallskip

\centerline { \sl Ann Arbor, MI 48109 U. S. A. }

\vskip2truecm

\centerline{\bf Abstract}
\bigskip
We perform a finite group
analysis on the quark mass matrices.  We argue that the
dominant terms should be proportional to class operators
of the group and that symmetry breaking to split the mass spectrum
and simultaneous diagonalizability to suppress flavor
changing neutral currents can be accomplished at this
point.  The natural setting is a multi-scalar model and
the scalar doublets can have masses of the weak
scale without any parameter tuning.   When we specialize
to $S_3$ as the group of choice,
we arrive at the results that the dominant mass terms are
$\lhook $democratic$\rhook $ and that the ratios of light masses
and the Cabbibo angle $\cong ({m_d\over m_s})^{1\over 2}$ are all
given by group parameters
in the breaking of $S_3$ to $S_2$.  A large mass
expansion is then performed and a generalized Wolfenstein
parameterization is given.  Further
breaking by way of introducing heavy-light transitions
in the down-type mass matrix is here related to the heavy-light
Cabbibo-Kobayashi-Maskawa elements.

\bigskip
\noindent
PACS number(s): 11.30.Hv,12.15.Ff,12.15.Cc,12.50.Ch
\vfill
\eject
\footline{\hss\tenrm\folio\hss}
\bigskip
One of the frontiers in understanding elementary interactions is the
organization of fermion masses, which in some effective way are
related to Yukawa couplings between fermions and scalars.  Many
proposals have been made and most are motivated by some conjectures
on physics at a much higher energy scale.  Typically, a certain
$\lhook $texture$\rhook $ is assumed for the Yukawa structure  and then a
renormalization
group analysis is performed to predict consequences for physical processes
which are currently experimentally reachable.  These are very ambitious
and formidable endeavours.
\bigskip
We shall take a different tack in the present discussion.  Our starting
point is to accept what we know from the data about fermion masses and
mixing between up and down sectors at the electroweak scale.  Several
features stand out: the almost decoupling of the top and bottom heavy
quarks from the lighter ones, the high degree of suppression of flavor
changing neutral currents at low energies, and the validity of the
Wolfenstein parameterization.  We then ask the question: How much of
this can be understood by applying symmetry considerations?  We argue
in this note that one can achieve quite a lot in this regard.  Of course,
some assumptions
need to be made along the way, and they will be explicitly stated.  They
have to do with symmetry breaking, which should be familiar to most of
us, drawing upon past experience.  We remark that this approach
may be complementary to the top down method just mentioned.  One advantage
here is an immediate link between physical parameters and those introduced
in the group analysis.
\bigskip
Before being specific, let us outline how such an analysis is developed.
Consider a group with a finite number of elements $g_i$.  We can partition
these elements into disjoint conjugate classes $C_j$.  Because $C_j$ commute
with each other and can be made hermitian, they are a part of the complete
set of observables and can be used to label states.$^{(1)}$  Also, because all
elements of the group commute with these class operators, $C_j$'s are
invariants.  As a zeroth order approximation, i.e., before symmetry
breaking is introduced, the
interaction which is responsible for mass
generation for either charged ${2\over 3}$ or
$-{1\over 3}$ type quarks is a linear combination of these class
operators, which we
write generically as
$$M_0=\sum a_j C_j. \eqno (1)$$
Because we are dealing with a finite group, the elements $g_i$ can
be made unitary, and the invariance under the proposed symmetry is
$$g_i M_0 g_i^{-1} =M_0.  \eqno (2)$$
\bigskip
The spectrum of $M_0$, which splits quarks into heavy and light
species, generally has some degeneracy at this level.  Past experience
leads us to speculate that the degeneracy is lifted by symmetry
breaking along some directon in the group space.  Thus, one assumes that
another term
$$M_1=\sum b_kg_k, \eqno (3)$$
accounts for that, where the sum is over a set of elements, such that
symmetry of some subgroup remains.  Therefore, $M_1$ must be expressible
as a function of the class operators of the subgroup.  This forces
conditions on b's, reducing their independent number.
\bigskip
We must digress at this point to discuss the problem of flavor changing
neutral currents.  As one follows the discussion so far, one must wonder
about the mechanisms which cause the division of $M$ into $M_0$ and $M_1$.
The current lore is that there may be different $SU(2)$ Higgs
doublets, which couple separately to $M_0$ and $M_1$.  We accept
this and will not be discussing the dynamical details pertaining to such
scalars at this juncture.  The
only issue we want to bring up is that if the scalars are distinct,
they will generally introduce tree level flavor changing neutral
current processes.$^{(2)}$  The reason is that if we write out the scalars
explicitly, we have
$$M(x)=\sum a_j'C_j\phi_0(x) + \sum b_k'g_k\phi_1(x), \eqno (4)$$
where the first and second terms on the right hand side, respectively,
come from $M_0$ and $M_1$.  Fermion masses are induced by replacing
the fields with their vacuum expectation values
$$\phi_{0,1}\rightarrow v_{0,1}, \eqno (5)$$
and performing a bi-unitary transformation $U^\dagger MV$.  Because of the
space-time dependence, such a transformation cannot diagonalize
$M(x) M(x)^\dagger$ for all x, unless
$$M_0 M_1, M_0 M_1^\dagger, M_1 M_1^\dagger,
\eqno (6)$$
commute.  We recall that $M_0=M_0^\dagger,$ and $[M_0,M_1]=0$ is automatic
by the very nature of $M_0$ being made of class operators.
Commutativity would be
trivial if $M_1=M_1^\dagger$ also.
However, in order to lift all
degeneracies at this point, hermiticity of $M_1$ may not be warranted and
commutativity should be checked.  If satisfied, then
under rather general scalar self interaction, the dominant part of the
induced flavor changing neutral
currents can in fact be avoided at least up to the one loop level.$^{(3)}$
We call the commutativity requirement radiatively natural.  The gist is
due to a result that the otherwise worrisome divergent
pieces of the one loop
contributions can be absorbed into wave function renormalizations
without spoiling simultaneous diagonalizability.
\bigskip
We have generated masses for the heavy quarks through $M_0$, and masses
for the light quarks and their mixing mostly through $M_1$.  The requirement
of simultaneous diagonalizability probably will not induce misalignment
between the heavy and the light states of the up and down type quarks if
we assume that the symmetry basis vectors in both sectors are the same;
i.e., the Cabibbo-Kobayashi-Maskawa (CKM) matrix
elements $V_{td,ts,cb,ub }$ vanish at this level.
If our picture is in concordance with nature, there must exist another
piece $M_2$, which gives rise to finite, albeit small,  heavy-light
mixing matrix elements, and which also results in flavor violation in
heavy-light transitions.  We shall now turn to an example to give some
specifics.
\bigskip
A finite group which is suggested empirically is the symmetric group
$S_3$,$^{(4)}$ with group elements $\{ e, (12),(13),(23), (123),(132)\}$,
where e is the identity, (12) is the operation of exchanging entries in
positions 1 and 2, and (123) corresponds to $1\rightarrow 2 \rightarrow 3
\rightarrow 1$, etc.  Let us take the up quark sector 3x3 mass matrix
$$\bar u_L M_u u_R, \eqno (7)$$
which we assume to be invariant under
$$\bar u_L \rightarrow \bar u_L g_i, \ \ u_R\rightarrow g_i^{-1}u_R,
\eqno (8)$$
for $g_i \epsilon S_3$.  The conjugate classes are $\{ e\},
\{ (12), (13),(23)\},$ and $\{(123),(132) \}$, with the
concomitant class operators
$$C_1=e, \ \ C_2=(12)+(13)+(23), \ \ C_3=(123)+(132). \eqno (9)$$
{}From the group table, one finds $C_3=(C_2)^2/3-C_1$, which
means that at most two of these class operators need be specified
to label states.
\bigskip
The three quark states are assumed to be linear combinations of the
basis vectors
$|\alpha , \alpha , \beta >, \ |\alpha ,  \beta, \alpha  >,$
and $|\beta, \alpha , \alpha  >$, on which the symmetry operations
act on the entries $\alpha $ and $\beta $, e. g.
$$\eqalign {(13)&( |\alpha , \alpha , \beta >, \ |\alpha ,  \beta, \alpha  >,
|\beta, \alpha , \alpha  >)\cr
&=(|\beta, \alpha , \alpha  >,| \alpha , \beta ,\alpha  >,
| \alpha , \alpha , \beta  >)\cr
&=(|\alpha , \alpha , \beta >, \ |\alpha ,  \beta, \alpha  >,
|\beta, \alpha , \alpha  >)\pmatrix {0&0&1\cr 0&1&0\cr
 1& 0& 0\cr},\cr}  \eqno (10)$$
from which one obtains the (reducible) matrix representation.  One can
easily show that on these states, the class operator
$$ C_2=\pmatrix { 1& 1& 1\cr  1& 1& 1\cr
 1& 1& 1\cr}, \eqno (11)$$
and $C_1+C_3=C_2$.  Looking at their eigenvalues, one sees that $C_2$
has (0,0,3), which makes it empirically rather compelling to take$^{(4)}$
$$M_0=m_0 C_2, \eqno (12)$$
to give mass to the top quark, where $m_0$ is a real constant carrying
the dimension of mass.
\bigskip
To account for the light quarks c and u, we {\it assume } that
$M_1$ is along some direction such that $S_2$ is the residual
symmetry.  For $S_2$, there are only two elements $\{e, g\}$, with
$g^2=e$.  To make this general, we write
$$M_1=m_1g, \ \ \ g=a_1e+a_2(12)+a_3(13)+a_4(23)+a_5(123)+a_6(132),
\eqno (13)$$
where $m_1 \ll m_0$ is another real constant with the dimension of mass.
A set of conditions which yield the requirement $g^2=e$ is
$$a_1=0, \ \ a_5+a_6=0, \ \ a_2+a_3+a_4=1, \ \ $$
and
$$a_2^2+a_3^2+a_4^2=1+2a_6^2 \eqno(14)$$
We shall make the choice that all the a's are real.  (This
results in a non-hermitian reducible g, which is what we need
to separate the light masses. The residual symmetry acts on the
mass matrix $M_0+M_1$, but not on the states.)  It is easy to verify
that the simultaneous diagonalizability conditions of Eq.(6) are
satisfied, basically because $M_0$ is unitarily equivalent to a
diagonal matrix with only one non-vanishing entry.  The eigenvalues
of $M_1M_1^\dagger$ are
$$\lambda ^2_{1,2}=m_1^2(1+6a_6^2\mp 2a_6 \sqrt {3+9a_6^2}),\ \ \
\Delta \lambda _3^2=m_1^2, \eqno (15)$$
which depend on $a_6$ only.  One can solve for it as
$$ a_6={m_c-m_u \over 2 \sqrt {3m_um_c}}. \eqno (16)$$
The corresponding eigenvectors are
$$|\lambda _{1,2}^0>=N_{1,2}|x_{1,2},\ y_{1,2},\ -(x_{1,2}+y_{1,2})>, \ \
|\lambda _3^0> ={1\over \sqrt 3}|1,\ 1,\ 1>, $$
with
$${y_{1,2}\over x_{1,2}}= {\mp \sqrt {3+9a_6^2}+3a_4-1 \over
3a_2-1}, \eqno (17)$$
and $N_{1,2}$ are normalization factors.
\bigskip
With the conditions of Eq.(14) and the a's being real, we have three
independent parameters, which may be chosen as $m_1, \ a_2 $
and $a_6$.  They uniquely give the masses $m_u\cong \lambda _1, \
m_c\cong \lambda _2$ and the relative weight $y/x$ of the physical
states $|\lambda _{1,2}^0>\cong |u,c>$.  We can replicate the same
analysis for the down sector and obtain similar results,
which we use primes to denote.
A further assumption of charge independence $a_2=a'_2$
reduces the number of parameters to five, which is in agreement with
the count of
$m_{c,u}, m_{s,d}$ and the Cabibbo angle
$sin\theta _c\cong V_{us}\cong  <\lambda _1^0|\lambda _2'^0>.$
\bigskip
A particular interesting case is when
$$a_2=a'_2=1,\eqno (18)$$
which gives, because of Eq.(14) with a choice of signs,
$$a_3=-a_6, \ a_4=a_6, \ a'_3=-a'_6, \ a'_4=a'_6. \eqno (19)$$
These lead to
$$sin \theta _c={({m_d\over m_s})^{1/2}-({m_u\over m_c})^{1/2}
\over (1+{m_d\over m_s})^{1/2}(1+{m_u\over m_c})^{1/2}}. \eqno (20)$$
As well-known, this is quite close to the measured value for the Cabibbo
angle.$^{(5)}$  The mixing angle $\theta _c$ is a dynamical signature in
the group space, pointing to that direction which seeks out the residual $S_2$
symmetry.  Although at this time
we have not been able to associate any deeper
meaning to this choice, other than the fact that
the values for $a_{2,3,4}$ look quite symmetrical, it
does illustrate succinctly the capability to relate to data.
\bigskip
We may wonder whether there is any freedom in introducing further
terms for the light sector.  In other words, is there a $\delta M$,
which is simultaneously diagonalizable with $M_1$ in the sense of Eq.(6)?
By using $g^2=e$, one can show that the only necessary condition is
$$\delta M M_1^\dagger M_1=M_1M_1^\dagger \delta M, \eqno (21)$$
which can be solved to give
$$\delta M=h_1C_2+h_2((123)-(132)), \eqno (22)$$
where $h_{1,2}$ are some arbitrary constants.  This matrix is also
simultaneously diagonalised with $M_0$ and
therefore does not lead to any CKM heavy light mixing.  Besides, there
is no underlying group argument as we had for $M_1$ to justify its
being. We shall just discard it.
\bigskip
To discuss the CKM heavy light mixing, it is convenient to make a
unitary transformation to decompose into the irreducible subspaces, viz.
$3 \rightarrow 1\oplus 2$.  This is done by
$$g_i \rightarrow {\cal U}^\dagger g_i {\cal U},$$
where
$${\cal U} =\pmatrix {
{-1\over \sqrt 6} &{-1\over \sqrt 2}&{1\over \sqrt 3}\cr
{-1\over \sqrt 6} &{1\over \sqrt 2}&{1\over \sqrt 3}\cr
{2\over \sqrt 6} &0&{1\over \sqrt 3}\cr}.\eqno (23)$$
Then, the mass matrix
$$M_0+M_1\rightarrow \pmatrix {(M_1)_{2\times 2}&0_{2\times 1}\cr
0_{1\times 2}& \bar m_0\cr}, $$
in which $\bar m_0=3m_0+m_1$ and
$$(M_1)_{2\times 2}
=m_1({\sqrt 3\over 2}(a_2-a_3)\sigma _1+\sqrt 3a_6i\sigma _2
+{1\over 2}(-a_2-a_3+2a_4)\sigma _3). \eqno (24)$$
\bigskip
We make the ansatz that heavy light transition is due to
$$M_2=\pmatrix {0&0& \Delta f_x \cr
0&0& \Delta f_y\cr
\Delta d_x&\Delta d_y&0\cr}, \eqno (25)$$
in which $\Delta $d's and $\Delta $f's are complex numbers of order at
most $m_1$, so that all low energy flavor changing neutral
processes due to the absorption, emission or exchange of attendant Higgs
scalars will be suppressed by heavy quark propagators.
\bigskip
We are now ready to complete our discussion of the CKM matrix
by performing an expansion in inverse powers of $m_b$ and $m_t$.$^{(6)}$
We note that for $M_u=M_0+\epsilon M_1 +\epsilon M_2$. we have
$$M_u M_u^\dagger =\bar m_0^2 \pmatrix {0&0&0\cr
0&0&0\cr 0&0&1\cr}+\epsilon \bar m_0\pmatrix {0&0&\Delta f_x\cr
0&0&\Delta f_y \cr \Delta f^\star _x&\Delta f^\star _y&0\cr}
+O(\epsilon ^2).  \eqno (26)$$
$\epsilon $ is a counting parameter in the inverse mass expansion,
which will be set to unity afterwards.  Note that because we
are dealing with left-left mixing, the second term on the
right hand side of the last equation, which is the only $O(\epsilon )$ term,
has dependence on $\Delta f$'s only.  $\Delta d$'s are not
measurable to this order.
\bigskip
It is a simple matter to solve for the eigenvectors to obtain
$$|u,c>=|\lambda _{1,2}>\cong |\lambda _{1,2}^0>-{\Delta F^\star_{1,2}
\over m_t} |\lambda _3^0>,$$
$$|t>=|\lambda_3>\cong |\lambda_3^0>+{\Delta F_1 \over m_t}|\lambda _1^0>
+{\Delta F_2 \over m_t}|\lambda _2^0>,$$
where
$$\Delta F_{1,2}\equiv <\lambda _{1,2}^0|{\cal U}\pmatrix {\Delta f_x\cr
\Delta f_y \cr 0\cr}=-N_{1,2}(\sqrt {3\over 2}(x+y)_{1,2}\Delta f_x
+\sqrt {1\over 2}(x-y)_{1,2}\Delta f_y).  \eqno (27)$$
{}From these, we form the CKM matrix elements
$$V_{ud}=<u|d>\cong <\lambda _1^0|\lambda_2'^0>=cos \theta _c,$$
$$V_{us}\cong sin \theta _c, \ \ \ V_{cd}\cong -sin \theta _c,
\ \ \ V_{cs}\cong cos \theta _c,$$
$$V_{td}\cong {\Delta F^\star _1 \over m_t}cos \theta _c
-{\Delta F^\star _2 \over m_t}sin \theta _c-{\Delta F^{'\star}_1\over m_b},$$
$$V_{ts}\cong {\Delta F^\star _1 \over m_t}sin \theta _c
+{\Delta F^\star _2 \over m_t}cos \theta _c-{\Delta F^{'\star}_2\over m_b},$$
$$V_{ub}\cong -V^\star_{td}cos\theta _c -V^\star _{ts}sin \theta _c,$$
$$V_{cb}\cong V^\star_{td}sin\theta _c -V^\star _{ts}cos \theta _c,$$
$$V_{tb}\cong 1.  \eqno (28)$$
These expressions have further corrections of order ${1\over m_b^2},
\ {1\over m_b m_t}, \ {1\over m_t^2}$.  Eqs.(28) may be taken as a slightly
generalized Wolfenstein parameterization.$^{(7)}$  If we assume
$\Delta F_{1,2}/m_t\ll \Delta F'_{1,2}/m_b$ and drop the former,
the number of parameters
we need to incorporate heavy-light transitions in CKM
matrix is three, namely the magnitudes of $\Delta f'_{x,y}$ and the
relative phase, which is precisely what we need to specify in
general.  CP violation is intimately tied up with flavor
violation in the heavy-light connection.
\bigskip
Because of simultaneous diagonalizability of $M_0$ and $M_1$, there
is no flavor changing neutral current due to tree level scalar
exchanges in the light sector.  The masses of those scalar doublets
associated with $M_0$ and $M_1$ can take on single
Higgs values $\sim m_W$ as in conventional Standard Model analysis.
Particularly, they will not
give rise to disproportionate surprises in $K^0$-$\bar K^0$ or
$D^0$-$\bar D^0$ systems.$^{(2)}$  New physics most likely will be first
revealed in processes through the intermediary of top and bottom
quarks, whence exploration in future B-factories should be most
interesting.
We are looking into phenomenological manifestation of the
terms $\Delta d$, $\Delta d'$, $\Delta f$, $\Delta f'$ and the
accompanying scalars.
\bigskip
In summary, we have argued that if the flavor space admits an
approximate symmetry of a finite group, then the dominant
piece of the Yukawa interactions should be a function of some class
operators of that group.  Ratios of light quark masses and the Cabbibo
angle are given by directional parameters of some subgroup into which the
original symmetry breaks.  The dynamical issue of masses and mixing is
then shifted into the eventual determination of these parameters from some
first principle.  $S_3$ is used to show explicitly
how this works.  We have been able to match the independent parameters in
the analysis to basically quark masses and CKM angles.  There is no flavor
changing neutral current, until the
last stage when heavy-light transition terms are introduced
to account for heavy-light CKM mixing.
\bigskip
This work has been partially supported by the U. S. Department
of Energy.
\vfill
\eject
\noindent
{\bf References:}
\smallskip
\noindent
(1) See, for example, J.-Q. Chen, {\it Group Representation
Theory for Physicists}, (World Scientific, Singapore, 1989).
\smallskip
\noindent
(2) T. P. Cheng and M. Sher, Phys. Rev. {\bf D35}, 3484 (1987);
L. Hall and S. Weinberg, Phys. Rev. {\bf D48}, R979 (1993).  These
works illustrate how attempts have been made to suppress flavor
changing neutral current processes by postulating certain
structure on the Yukawa couplings.  The Higgs scalars typically
have masses $\sim  1 \ Tev$.
\smallskip
\noindent
(3) J. M. Frere and Y.-P. Yao, Phys. Rev. Lett. {\bf 55}, 2386
(1985).
\smallskip
\noindent
(4) H. Harari, H. Haut and J. Weyer. Phys. Lett. {\bf 78B},
459 (1978); Y. Chikashige, G. Gelmini, R. P. Peccei and M.
Roncadelli, Phys. Lett. {\bf 94B}, 499 (1980); C. Jarlskog,
{\it Proc. of the Internaional Symposium on Production and Decay of
Heavy Flavors}, Heidelberg, Germany, (1986); P. Kaus and S. Meshkov,
Mod. Phys. Lett. {\bf A3}, 1251 (1988), and {\it ibid} {\bf A4}, 603 (1989);
G. C. Branco, J. I. Silva-Marcos and M. N. Rebelo, Phys. Lett.
{\bf B237}, 446 (1990); Y. Koide, Z. Phys. {\bf C45}, 39 (1989);
H. Fritzsch and J. Plankl, Phys. Lett. {\bf B 237} 451 (1990).
\smallskip
\noindent
(5) S. Weinberg, {\it Transactions of the New York Academy of Sciences},
Series II, Vol.{\bf 38}, 185 (1977); H. Fritzsch, Phys. Lett. {\bf 70B},
436 (1977).
\smallskip
\noindent
(6) Y.-P. Yao, Phys. Rev. {\bf D51}, 5240 (1995).
\smallskip
\noindent
(7) L. Wolfenstein, Phys. Rev. Lett. {\bf 51}, 1945 (1983).

\end